\documentclass[useAMS,usenatbib,usegraphicx,twocolumn]{mn2e}

\usepackage{epsfig}
\usepackage{times}
\usepackage{natbib}
\usepackage{color}
\usepackage{multirow}
\usepackage{verbatim}
\usepackage[figuresright]{rotating}
\usepackage{afterpage}  
\usepackage{lipsum}
\usepackage{comment}
\usepackage{mathptmx}
\usepackage{amsmath}
\usepackage{amssymb}
\usepackage{url}
\usepackage{refcount}
\def\lsim{\lower.5ex\hbox{$\; \buildrel < \over \sim \;$}}
\def\gsim{\lower.5ex\hbox{$\; \buildrel > \over \sim \;$}}

\def\apj{ApJ}
\def\apjs{ApJS}
\def\araa{ARAA}
\def\mnras{MNRAS}

\def\nar{New Astronomy Reviews}
\def\aj{AJ}
\def\apjl{ApJL}

\def\aap{Astronomy \& Astrophysics}

\newcommand\altaffilmark[1]{$^{#1}$}
\newcommand\altaffiltext[1]{$^{#1}$}

\title[Formation of Globular Cluster Candidates]{Formation of Globular Cluster Candidates in Merging Proto-galaxies at High Redshift: A View from the FIRE Cosmological Simulations
\vspace{-0.3cm}}

\vspace{-0.2cm}

\author[J. Kim et al.]{
\parbox[t]{\textwidth}{ 
Ji-hoon Kim\altaffilmark{1, 2, 3, 4}\thanks{E-mail: me@jihoonkim.org}, 
Xiangcheng Ma\altaffilmark{3}, 
Michael Y. Grudi\'{c}\altaffilmark{3}, 
Philip F. Hopkins\altaffilmark{3}, 
Christopher C. Hayward\altaffilmark{5,6}, 
Andrew Wetzel\altaffilmark{3,7,8,9}, 
Claude-Andr\'{e} Faucher-Gigu\`{e}re\altaffilmark{10}, 
Du\v{s}an Kere\v{s}\altaffilmark{11}, 
Shea Garrison-Kimmel\altaffilmark{3,4}, 
Norman Murray\altaffilmark{12,13}
}
\vspace*{6pt} \\
\altaffiltext{1}{Kavli Institute for Particle Astrophysics and Cosmology, SLAC National Accelerator Laboratory, Menlo Park, CA 94025, USA} \\
\altaffiltext{2}{Department of Physics, Stanford University, Stanford, CA 94305, USA} \\
\altaffiltext{3}{TAPIR, Department of Astronomy, California Institute of Technology, Pasadena, CA 91125, USA} \\
\altaffiltext{4}{Einstein Fellow} \\ 
\altaffiltext{5}{Center for Computational Astrophysics, Flatiron Institute, New York, NY 10010, USA} \\ 
\altaffiltext{6}{Harvard--Smithsonian Center for Astrophysics, Cambridge, MA 02138, USA} \\
\altaffiltext{7}{Carnegie Observatories, Pasadena, CA 91125, USA} \\
\altaffiltext{8}{Department of Physics, University of California, Davis, CA 95616, USA}\\
\altaffiltext{9}{Caltech-Carnegie Fellow} \\ 
\altaffiltext{10}{CIERA, Department of Physics and Astronomy, Northwestern University, Evanston, IL 60208, USA}\\
\altaffiltext{11}{Department of Physics, Center for Astrophysics and Space Sciences, University of California at San Diego, La Jolla, CA 92093, USA}\\
\altaffiltext{12}{Canadian Institute for Theoretical Astrophysics, University of Toronto, Toronto, ON M5S 3H8, Canada}\\ 
\altaffiltext{13}{Canada Research Chair in Astrophysics} 
\vspace{-0.5cm}
}

\begin{document}

\date{Submitted April 2017}

\pagerange{\pageref{firstpage}--\pageref{lastpage}} \pubyear{2017}

\maketitle

\label{firstpage}

\begin{abstract}

Using a state-of-the-art cosmological simulation of merging proto-galaxies at high redshift from the {\it FIRE} project, with explicit treatments of star formation and stellar feedback in the interstellar medium, we investigate the formation of star clusters and examine one of the formation hypothesis of present-day metal-poor globular clusters.  
We find that frequent mergers in high-redshift proto-galaxies could provide a fertile environment to produce long-lasting bound star clusters.   
The violent merger event disturbs the gravitational potential and  pushes a large gas mass of $\gtrsim 10^{5-6}\,{\rm M}_{\odot}$ collectively to high density, at which point it rapidly turns into stars before stellar feedback can stop star formation.
The high dynamic range of the reported simulation is critical in realizing such dense star-forming clouds with a small dynamical timescale, $t_{\rm ff} \lesssim 3$ Myr, shorter than most stellar feedback timescales. 
Our simulation then allows us to trace how clusters could become virialized and tightly-bound to survive for up to $\sim$420 Myr till the end of the simulation.  
Because the cluster's tightly-bound core was formed in one short burst, and the nearby older stars originally grouped with the cluster tend to be preferentially removed, at the end of the simulation the cluster has a small age spread. 
\end{abstract}


\section{INTRODUCTION}\label{intro}

By the interaction of gravity and pressure, gas becomes unstable and collapses to turn into stars.  
Observationally we know that most stars form in ``clustered fashion'', with some stars forming {\it en masse} in dense molecular clouds of $\gtrsim10^{4}\, {\rm M}_{\odot}$ and others in relatively loose associations (for reviews, see \citealt{2010ARA&A..48..431P} and \citealt{2010MNRAS.409L..54B}).
The fraction of stars that form in bound clusters, often referred to as $\Gamma$, is affected by the host galaxy environment, with galaxies with higher star formation activity such as starbursts having higher $\Gamma$'s than local spiral and dwarf galaxies \citep[e.g.,][]{2012MNRAS.426.3008K, 2015arXiv151108212A}.
Simulations of an isolated star clusters in an idealized setup have a long history and can now resolve the formation of individual stars (for recent reviews,  see \citealt{2013arXiv1304.4600K} and \citealt{2015NewAR..68....1D}).  
However, they cannot be used to study the formation of a dense environment in a galactic context, in which those star clusters spawn.  
It is also difficult to utilize this type of calculations in studying how bound versus unbound star clusters form, and how they evolve and survive in a galactic environment.  
A  high-resolution, galaxy-wide simulation is required to address these questions.

However, star clusters have rarely been simulated in a galactic context, with numerical accuracy high enough to resolve their dynamical evolution.  
Note that in most previous galaxy simulations the mass of each resolution element approximately equates to that of a star cluster.
Only after we reach a mass resolution of $\lesssim10^3\, {\rm M}_{\odot}$ can we begin to reliably resolve the kinematics amongst the member star particles of a massive $\sim10^{5-6}\, {\rm M}_{\odot}$ cluster in a galaxy-scale simulation.\footnote{We caution that even a simulation with  $\sim10^3\, {\rm M}_{\odot}$ resolution cannot reproduce the phase-space of {\it actual stars} in a cluster.  See Section \ref{results-evolution} for more discussion on numerical resolution and the ``intra-cluster'' evolution. \label{resolving-actual-stars}}$^{,}$\footnote{Readers should also note that the target cluster mass of $\sim10^{5-6}\, {\rm M}_{\odot}$ which even the highest-resolution simulations aim to resolve is still close to the massive end of the cluster mass function \citep[that could reach as low as $\sim10^{2-3}\, {\rm M}_{\odot}$;][]{2009A&A...494..539L, 2010ARA&A..48..431P, 2014CQGra..31x4006K}.}
It is also required that realistic physics models be in place which describe star formation and stellar feedback at such high resolution.  
Most previous galaxy-scale simulations lack such sophisticated physics models.

Recently, however, this is becoming increasingly possible with modern cosmological simulations. 
In this paper, we use simulations from the {\it FIRE} project \citep[{\it Feedback In Realistic Environments};][]{2014MNRAS.445..581H}\footnote{The website is http://fire.northwestern.edu/.\label{fire-website}} to study how star clusters form and evolve in high-redshift merging proto-galaxies. 
This has been proposed \citep{1995AJ....109..960W, 1996AJ....112.1839S, 2006astro.ph..6036S} as a formation channel for the present-day old, metal-poor ``blue'' globular clusters (GCs) which tend to be distributed throughout the galactic halo (as opposed to metal-rich ``red'' GCs which are found mostly in galactic bulges). 
For reviews of their properties and various hypotheses of their origin, see \citet{1968ApJ...154..891P, 1979ARA&A..17..241H, 1985ApJ...298...18F, 2001stcl.conf..223H, 2006ARA&A..44..193B}. 
Our investigation is complementary to other recent studies which have explored alternative GC formation scenarios: for example, in binary merging galaxies at low redshift \citep[e.g.,][]{2004ApJ...614L..29L, 2008MNRAS.389L...8B, 2011MNRAS.414.1339K, 2012MNRAS.421.1927K, 2015MNRAS.446.2038R, 2017ApJ...844..108M}, or in primordial mini-halos at redshift $z \gtrsim10$ \citep[e.g.,][]{2016ApJ...823...52K, 2016ApJ...831..204R}. 

Specifically, we carry out an investigation using a cosmological ``zoom-in'' simulation from \citet{2015MNRAS.453..960M} at $z\sim6$, with $\sim$\,pc-scale resolution, run with the {\it FIRE} physics that includes star formation {\it only} in dense, self-gravitating gas and stellar feedback from supernovae (SNe), stellar winds, photoionization and photoelectric heating, and radiation pressure. 
Our setup allows us to explicitly follow the formation and evolution of at least the most massive star clusters. 
We show that, in the simulation, most stars form in unbound associations but some form in resolved, bound clusters. 
Frequent mergers in high-redshift proto-galaxies provide a fertile environment to produce the latter population by pushing large gas masses ($\gtrsim 10^{5-6}\, {\rm M}_{\odot}$) collectively to high density, at which point it turns into stars before stellar feedback can disrupt the clouds. 
We explore subsequent dynamical evolution, virialization, and the age and metallicity spreads of the resulting clusters. 

The remainder of this article is organized as follows.  
In Section \ref{methodology} we detail the simulation code and methods.  
Section \ref{results} presents results of the simulation focusing on the formation and evolution of bound clusters. 
Finally in Section \ref{conclusion} we summarize our findings and conclusions.

\section{METHODOLOGY}\label{methodology}

The simulation studied in this work is one of a suite of high-resolution cosmological simulations at high redshift from the {\it FIRE} project, and is presented and described in detail in \citet[][{\it z5m10h} run therein]{2015MNRAS.453..960M}. 
We briefly review its important features for completeness. 
The simulation was run using the {\sc Gizmo} code \citep{hopkins2015}\footnote{The website is http://www.tapir.caltech.edu/$\sim$phopkins/Site/GIZMO.html.} which solves gravity using a tree-particle mesh (TreePM) method with fully adaptive gravitational softenings (scaled to the inter-particle separation), and solves the hydrodynamics using the Lagrangian ``pressure-energy'' ({\sc P-sph}) formulation of smoothed particle hydrodynamics with various improvements to alleviate known issues with fluid mixing and shock-capturing in older SPH formulations \citep[see][]{hopkins2013, 2014MNRAS.445..581H}. 
The simulation follows a high-resolution Lagrangian region around a ``target halo'' of virial mass $\sim 10^{10}\,{\rm M}_{\odot}$ (stellar mass of the ``target galaxy'' $\sim 4.7\times10^{7}\,{\rm M}_{\odot}$) at $z \sim 6$, with fixed mass resolution 800 $h^{-1}{\rm M}_{\odot}$ and minimum force resolution 1.4 $h^{-1}\,{\rm pc}$ (proper) for gas (and fixed to 1.4 $h^{-1}\,{\rm pc}$ for star particles). 
This resolution is high enough to resolve the dynamics between member particles of a relatively massive star cluster studied in this paper (but see also footnote \getrefnumber{resolving-actual-stars}).  
At $z=5$ of the original \citet{2015MNRAS.453..960M} simulation, we identify a group of long-lasting bound clusters, one of which we hereafter call the cluster ``A'' (see Figures \ref{fig:gas_star}-\ref{fig:composite_2kpc}).
We specifically re-ran the interval from $z=7$ to $z=5$ of the original simulation to generate more snapshots at finer intervals during $\sim$80 Myr before and after the cluster ``A'' formed, in order to study when, where and how the cluster forms.\footnote{For the original \citet{2015MNRAS.453..960M}, 31 snapshots were produced at intervals coarsely spaced between $z = 7$ and $z = 5$; that is, $\Delta\, z = 0.1$ ($\Delta\, t_{\rm out} \sim 17$ Myr) for $6< z <7$, and $\Delta\, z = 0.05$ ($\Delta\, t_{\rm out} \sim 13$ Myr) for $5< z <6$.  For the presented re-run, we output 76 snapshots at intervals more finely spaced between $z = 7$ and $z = 5$, especially around the time of the cluster ``A'' formation; that is, $\Delta\, z = 0.01$ ($\Delta\, t_{\rm out} \sim 1.6$ Myr) for $6.5< z <7$, $\Delta\, z = 0.1$ for $6< z <6.5$, and $\Delta\, z = 0.05$ for $5< z <6$. \label{output-strategy}}  

Baryonic physics is treated using the {\it FIRE-1} model, described in detail in \citet{2014MNRAS.445..581H}.
Briefly, radiative heating and cooling in $10-10^{10}$ K take into account molecular, atomic, ionized, and metal-line processes (with 11 independently-tracked species), and includes photoheating by local sources and redshift-dependent ultraviolet background \citep{Faucher09}, and self-shielding. 
Star formation occurs via spawning of star particles {\it only} from gas which meets a series of criteria: it must be locally self-gravitating and Jeans-unstable according to a sink particle criterion (including both thermal and turbulent support; see \citealt{2013MNRAS.432.2647H}), molecular (following \citealt{2011ApJ...729...36K}), and denser than a star formation threshold $n_{\rm th} = 500\,\,{\rm cm}^{-3}$.\footnote{Note that \cite{2015MNRAS.453..960M} mistakenly stated $n_{\rm th} =  1000\,\, {\rm cm}^{-3}$ for the  {\it z5m10h} run.  The correct value is $1000\,h^2 \simeq 500\,\, {\rm cm}^{-3}$. \label{ma-threshold}} 
A new star particle inherits the mass, metallicity, gravitational softening length, and particle ID number from its progenitor gas particle.  
Once a star particle forms, the simulation explicitly tracks feedback from (1) local and long-range radiation pressure (including single-scattering and multiple-scattering of re-radiated infrared photons), (2) energy, momentum, mass and metal injection from SNe (Types Ia and II) and stellar mass-loss (OB and AGB-star winds), and (3) photoionization and photoelectric heating. 
The rates for each channel are calculated as a function of the stars' age and metallicity using a stellar population synthesis model {\sc Starburst99} \citep{1999ApJS..123....3L} assuming the \cite{2002Sci...295...82K} initial mass function (IMF). 

For more details on how each item above is implemented, we refer the interested readers to Section 2 of \cite{2012MNRAS.421.3488H}, and Section 3 and Appendix A of \cite{2014MNRAS.445..581H}.  
A series of {\it FIRE} simulations using cosmological and isolated initial conditions have reported reasonable star formation histories, stellar mass-halo mass relation, Kennicutt-Schmidt relation \citep{2014MNRAS.445..581H}, mass-metallicity relation \citep{2016MNRAS.456.2140M}, multiphase interstellar medium \citep[ISM;][]{2011MNRAS.417..950H}, galactic outflows \citep{2015MNRAS.454.2691M}, dense neutral hydrogen content of galactic halos \citep{2015MNRAS.449..987F, 2016MNRAS.461L..32F}, galaxy structures and metallicity gradients \citep{2017MNRAS.467.2430M}, among others. 
These studies validate the ``realistic'' baryonic physics we adopt in the present study.

\section{RESULTS}\label{results}

We start by giving an overview of the simulation result and the two distinct populations of star clusters. 
Then we discuss the formation, evolution, and composition of long-lived bound star clusters.

\subsection{Overview of Simulation Results: From The Host Galaxy To The Target Cluster}\label{results-overview}

\begin{figure}
\begin{center}
\hspace*{-0.1cm}\includegraphics[width=0.48\textwidth]{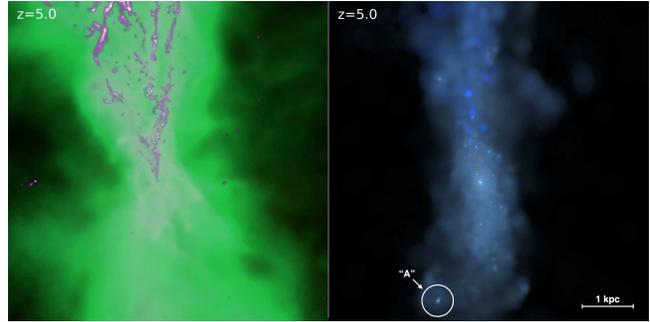}
    \caption{Snapshots of the target galaxy at $z=5$, and the star cluster of interest (cluster ``A'') studied here.  {\it Left:} gas surface density in a 6 kpc box (proper) centered on the target galaxy, where the brightness indicates projected gas density in a logarithmic scale and the colors encode different temperature ranges (e.g., magenta/white shows $T < 10^3$ K cold molecular gas, green $10^4 < T < 10^5$ K warm ionized gas).  {\it Right:} mock {\it ugr}-composite of stars in the same 6 kpc box.  Most stars are distributed in diffuse form, but several dense clusters are visually obvious within the galaxy, including the cluster ``A'' ({\it white circle}). 
\label{fig:gas_star}}
\end{center}
\end{figure}

Figure \ref{fig:gas_star} shows the $z=5$ snapshot of the target galaxy and the star cluster of interest studied in the present paper (cluster ``A'' defined in Section \ref{methodology}).  
Images are rendered as in the way described in \cite{2005ApJ...625L..71H, 2014MNRAS.445..581H}.  
In the left panel, molecular cloud complexes in cold filaments in the upper side of the panel indicate fresh flows of dense gas that has not turned into stars yet.   
The right panel shows multiple star clusters of masses ranging from $10^5$ to $10^7 \,\,{\rm M}_{\odot}$ scattered around the galaxy. 
Each of these clusters are dynamically resolved with $10^2 - 10^3$ star particles within $10 - 10^2$ pc half-mass radii (Figure \ref{fig:M_vs_Rh}).  
Among them, marked with a white circle of radius 300 pc is the cluster ``A''  that forms at $z = 6.92$ and survives $\sim$420 Myr afterwards.  

\begin{figure*}
\hspace*{-0.82cm}\includegraphics[width=1.115\textwidth]{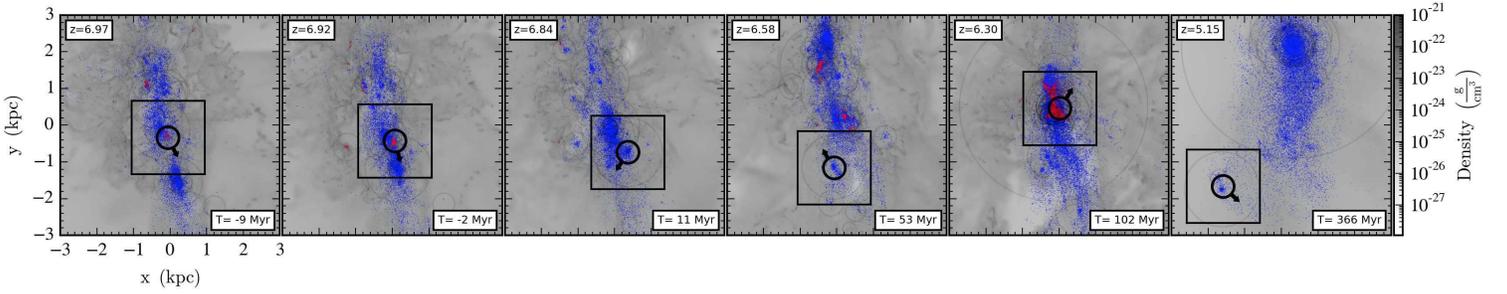}
    \caption{Distribution of star particles in a 6 kpc box (proper), evolving in time from $z= 6.97$ {\it (left)} to $z=5.15$ {\it(right)}.  Young star particles that are less than 5 Myr old are colored {\it red}, while the rest of the star particles are colored {\it blue}.  The gray background features the  density-weighted projection of gas density.  Marked with a {\it black circle} of radius 300 pc in each panel is the star cluster ``A'' (or its ``gas progenitor'' in the first and second panels) with an arrow indicating the direction of its movement.  The 2 kpc $\times$ 2 kpc square around the cluster corresponds to the region shown in Figure \ref{fig:composite_2kpc}.  Shown in the bottom right in each panel is the timestamp where 0 Myr corresponds to the moment the cluster ``A'' forms.  Cluster ``A'' forms in a proto-galaxy undergoing multiple rapid mergers.
\label{fig:star_6kpc}}
\vspace{5mm}
\end{figure*}

\begin{figure*}
    \hspace*{-1cm}\includegraphics[width=1.125\textwidth]{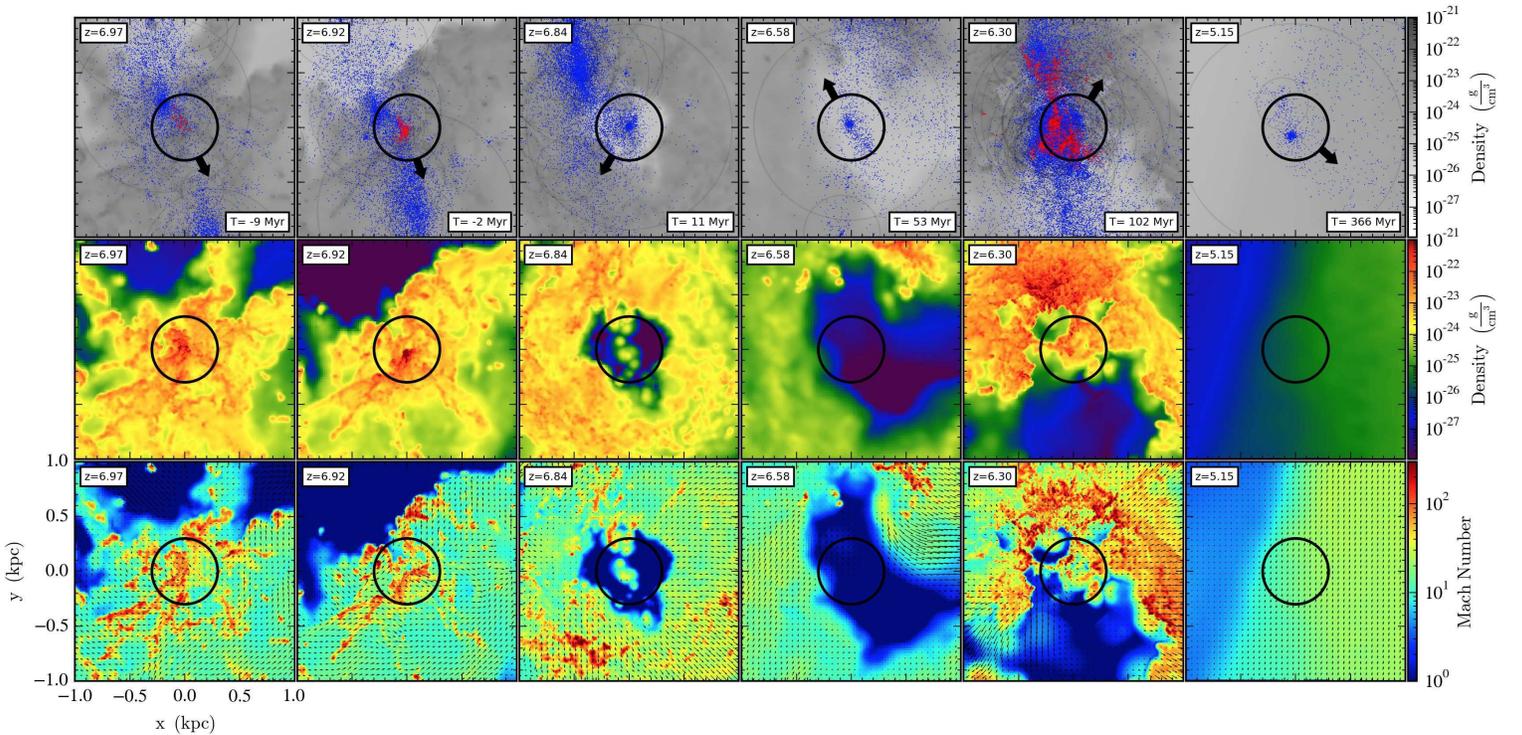}
    \caption{{\it Top:}  
    Same as Figure \ref{fig:star_6kpc} but in a 2 kpc box (proper) always centered on the cluster ``A'' (or its gas progenitor in the first and second columns from the left), evolving in time from $z= 6.97$ {\it (left)} to $z=5.15$ {\it(right)}.  This is a zoomed-in version of Figure \ref{fig:star_6kpc} (the square region) in the same style.  A {\it black circle} of radius 300 pc is centered on the cluster member particles.    
    {\it Middle:} 
    gas surface density in the same region.  
    {\it Bottom:}
    density-weighted projection of gas Mach number.
    An extremely dense cloud forms in a Mach $\sim$ 100 convergent flow (second column), then rapidly turns into stars within a couple of Myr.  The remaining gas is expelled over the next $\sim5-10$ Myr, leaving behind a dense, bound star cluster which persists as long as we run the simulation ($\sim$420 Myr).  
    \label{fig:composite_2kpc}}
\end{figure*}

\begin{figure*}
\begin{center}
\includegraphics[width=0.92\textwidth]{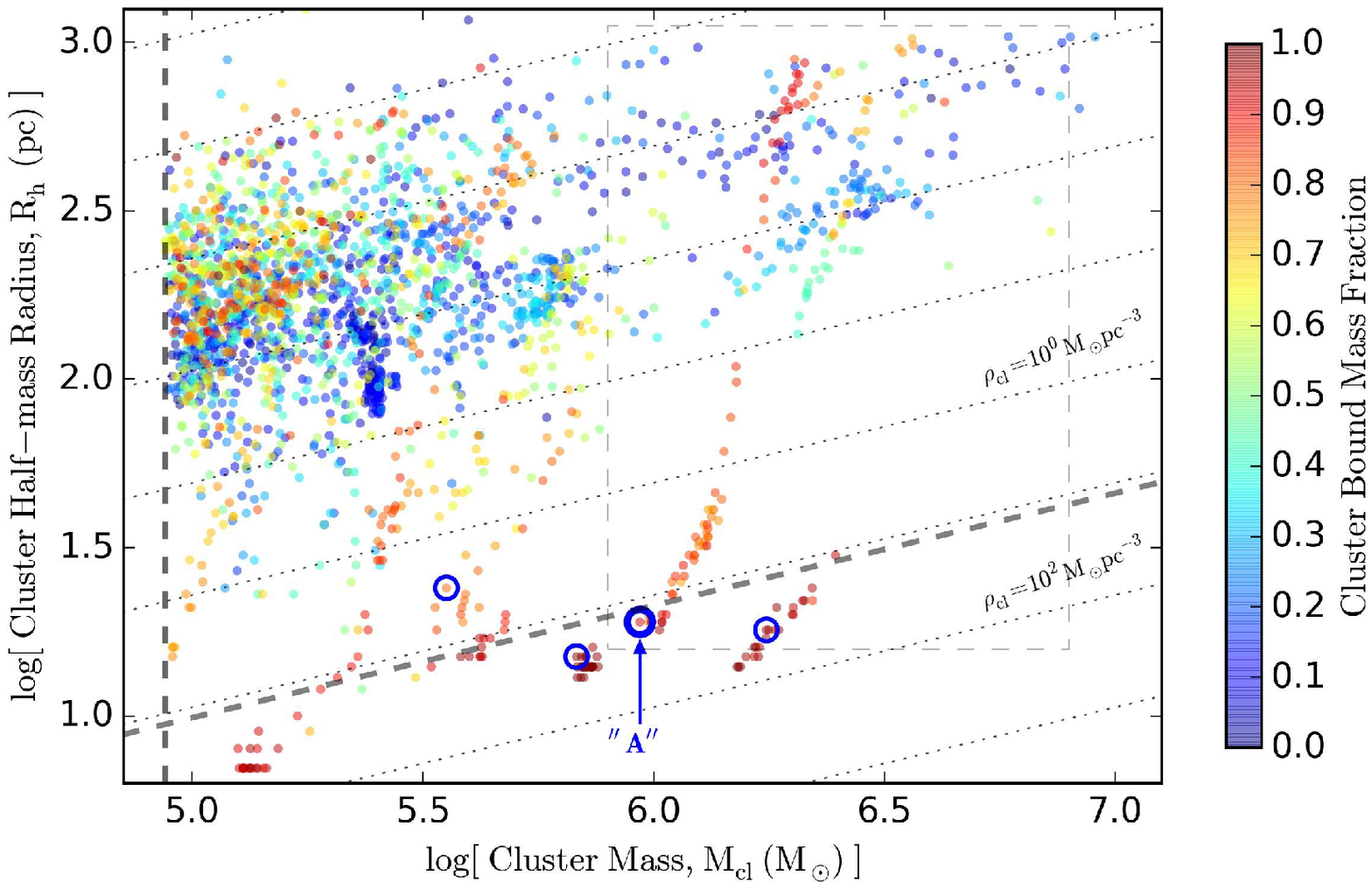}
    \caption{Cluster mass, $M_{\rm cl}$, versus half-mass radius, $R_{\rm h}$ for all star clusters identified by {\sc Rockstar} in all 76 snapshots finely spaced between $z=7$ and $z=5$,  that survived at least one output interval (e.g., $\Delta\, t_{\rm out} \sim 1.6$ Myr at $6.5< z <7$; see footnote \ref{output-strategy}).  Each data point is colored by the fraction of member particle masses that are bound to the cluster center, $f_{\rm bd}$.  The cluster ``A'' from Figures \ref{fig:gas_star}-\ref{fig:composite_2kpc} is marked with a {\it thick blue circle} at $z=5$.  Three other {\it thin blue circles} indicate the other long-lasting bound clusters identified at $z=5$.   For references, the thin dotted diagonal lines denote the slope of constant stellar density, while the thick dashed diagonal line marks the star formation threshold density, $500\,\, {\rm cm}^{-3}$ or $12\,\, {\rm M}_{\odot} \,{\rm pc}^{-3}$.  The thick dashed vertical line represents the threshold mass to be identified as a star cluster, $8.74 \times 10^4 \,\,{\rm M}_{\odot}$.  Most of the objects are young, loose associations with stellar densities $\sim 10^{-2}\,\, {\rm M}_{\odot} \,{\rm pc}^{-3}$ and little bound mass (upper left corner).  The few low-density associations with high bound mass fraction represent systems with a bound core and a loosely-bound ``envelope'' (upper right corner).  They rapidly evolve to become more dense clusters (blue circles; see Figure \ref{fig:M_vs_Rh_zoom-in} for more information that zooms in on the gray dashed rectangular region).  
\label{fig:M_vs_Rh}}
\end{center}
\end{figure*}

Figure \ref{fig:star_6kpc} reveals the movement of the cluster ``A'' throughout its host galaxy.\footnote{In all subsequent analyses we utilize the {\tt yt} toolkit \citep[][http://www.yt-project.org/, changeset d7f213e1752e]{yt}, a code-independent analysis platform adopted by the {\it AGORA} Initiative \citep{2014ApJS..210...14K_short, 2016ApJ...833..202K_short}.  To visualize fluid quantities such as gas density in Figures \ref{fig:star_6kpc}-\ref{fig:composite_2kpc} we employ {\tt yt}'s in-memory octree to which gas particles are assigned in a scatter step using the particles' hydrodynamic smoothing kernels ({\tt yt} parameters ${\tt n\_ref} = 4$ and ${\tt over\_refine\_factor}=2$).}$^{,}$\footnote{We refer the readers to Section \ref{results-population} for how star clusters including the cluster ``A'' are identified with {\sc Rockstar}, and to Section \ref{results-formation} for exactly how the cluster's ``formation time'' is defined.}
The second column captures the moment right before most star particles in the cluster ``A'' form (red, young star particles inside the black circle) during a major proto-galaxy merger.  
After the cluster forms, it moves to an orbit with a relatively large radius, piercing through the galactic nucleus at $z=6.30$ (fifth column), but later reaching far out into the extended galactic halo at $z=5.15$ (sixth column). 

To closely inspect the formation site of the cluster ``A'', in Figure \ref{fig:composite_2kpc} we zoom in on the 2 kpc $\times$ 2 kpc region around the cluster ``A''.
Before the cluster ``A'' forms, we see a dense gas clump forming in a galactic merger event (first and second columns; Section \ref{results-formation}). 
This efficiently forms new star particles in a short time, which then violently disperse the gas via stellar feedback, but only after a large number of star particles have formed (third and fourth columns). 
Later, we see the cluster pass through the center of the host galaxy, triggering a mass loss of the cluster (fifth column; Section \ref{results-evolution}).

\subsection{Two Populations of Star Clusters}\label{results-population}

We now examine the population of star clusters formed in our simulation between $z=7$ and $z=5$.  
To identify star clusters in the simulated galaxy, we utilize the 6-dimensional phase-space halo finder {\sc Rockstar} \citep{2013ApJ...762..109B}\footnote{The website is https://bitbucket.org/gfcstanford/rockstar/.} modified to run on star particles instead of dark matter particles.\footnote{For {\sc Rockstar} to work with its embedding {\tt yt}, we assume that all star particles have a constant mass, $874 \,\,{\rm M}_{\odot}$, the mean of all star particles.  In practice, the actual mass difference is less than a few tens of percent.}
To ensure that the identified star clusters are dynamically resolved and are not numerical artifacts, we only consider clusters with more than 100 member particles (or $> 8.74 \times 10^4 \,\,{\rm M}_{\odot}$).
We then directly extract the stellar mass, $M_{\rm cl}$, and spherical half-mass radius, $R_{\rm h}$, of each cluster.\footnote{$M_{\rm cl}$ is computed by summing up the real masses of member particles, not the constant masses we assumed for {\sc Rockstar}.  $M_{\rm cl}$ is thus slightly different from the group mass $M_{\rm vir,\,cl}$ found in the standard {\sc Rockstar} output.  Clusters are not necessarily a relaxed structure, but the two masses are different by less than a few tens of percent particularly if bound (justifying our use of $M_{\rm vir,\,cl}$ to calculate the bound mass fraction $f_{\rm bd}$). \label{rockstar-mvir}} 

In Figure \ref{fig:M_vs_Rh} we plot the $M_{\rm cl} - R_{\rm h}$ relation of all star clusters identified in 76 snapshots finely spaced between $z=7$ and $z=5$.  
We plot only those that survived at least one output interval (e.g., $\Delta\, t_{\rm out} \sim 1.6$ Myr at $6.5< z <7$; see footnote \ref{output-strategy}).  
In other words, we plot clusters that were found in two or more snapshots to eliminate spurious transient associations.  
Each data point is then colored by the fraction of member particle masses that are bound to the cluster, i.e., bound mass fraction $f_{\rm bd}$ derived from the standard {\sc Rockstar} output (but see footnote \getrefnumber{rockstar-mvir}).  
Additionally, four star clusters of mass above $10^{5.5}\,\,{\rm M}_{\odot}$ that survived more than 300 Myr are annotated with blue circles. 
These {\it long-lasting, bound star clusters} are identified at $z=5$ by considering the mass $M_{\rm cl}$, bound mass fraction $f_{\rm bd}$, and lineage information (see Section \ref{results-evolution}).
These clusters are grouped with few gas particles and almost no dark matter particles, thus not associated with dark matter over-densities or galactic subhalos.  
Among them, the cluster ``A'', studied extensively here, is marked with a thicker blue circle at $z=5$.     
The force resolution of the simulation is well below the $y$-axis range (see Section \ref{methodology}), indicating that clusters shown here are not affected by force softening, although they may of course be affected by the finite particle number (i.e., mass resolution).  

In this figure various features are prominent.  
First, a large fraction of the clusters congregate around $M_{\rm cl} \sim 10^{5.0-5.5}\, {\rm M}_{\odot}$ and $R_{\rm h} \sim 10^{1.5-2.5}\,{\rm pc}$, giving mean stellar densities, defined as $\rho_{\rm cl} \equiv 3M_{\rm cl}/(8\pi R_{\rm h}^3)$, of $\sim 10^{-2}\,\,{\rm M}_{\odot} \,{\rm pc}^{-3}$.
The majority of these clusters are loose ``unbound associations'' with low $f_{\rm bd}$. 
These associations, according to our lineage tree analysis (Section \ref{results-evolution}), often dissolve into an extended stellar disk or bulge in $10-30$ Myr by rapidly losing their unbound member particles.
This behavior is expected since most of the unbound associations inherit the properties of {\it normal} molecular clouds that are only marginally gravitationally bound after turning just a few percent of the cloud mass into stars.
Indeed, their observational counterparts such as OB associations tend to drift apart in a $\gtrsim10$ Myr timescale.
Readers should note that the density of a cluster/association (or its gas progenitor) can be much less than the star formation threshold ($500\,\, {\rm cm}^{-3}$ or $12\,\,{\rm M}_{\odot} \,{\rm pc}^{-3}$), because stars form in a small self-gravitating sub-clumps above the threshold density, distributed widely over a lower-mean-density giant molecular cloud (GMC) complex.  

A much smaller fraction of the stellar mass is represented by self-gravitating bound clusters that often survive for a longer time.  
For example, the long-lasting bound clusters identified at $z=5$ (blue circles in Figure \ref{fig:M_vs_Rh} and their ancestors) have radii as small as  $R_{\rm h} \sim 10$ pc with mean stellar densities as large as $\sim 10^{2}\,\,{\rm M}_{\odot} \,{\rm pc}^{-3}$. 
Their mean densities are often comparable to or larger than the star formation threshold, and their typical bound mass fractions are higher than 80\% (red face color).  
In addition, they tend to maintain their locations on the $M_{\rm cl} - R_{\rm h}$ plane for $\gtrsim 100$ Myr (as indicated by groups of red points crowded around the four blue circles).  
This second group of star clusters could be categorized as ``bound clusters'', which could potentially become candidates for present-day metal-poor ``blue'' GCs.  
We note that these numerically formed clusters are similar in their sizes and masses to the objects observed at $z\gtrsim6$ by  \cite{2017MNRAS.467.4304V} and conjectured as proto-GCs.

\begin{figure}
\begin{center}
\includegraphics[width=0.48\textwidth]{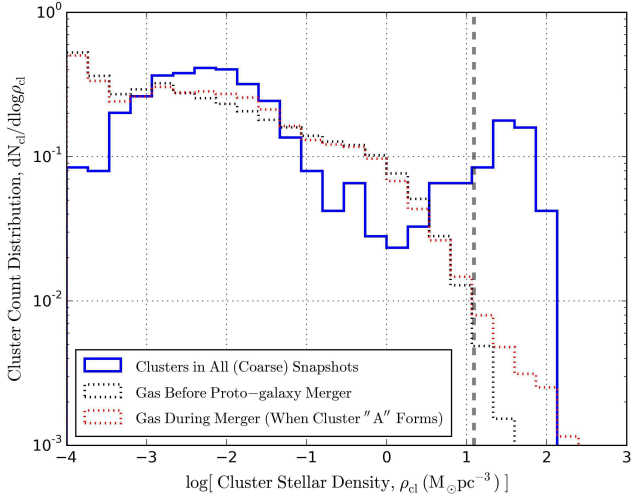}
    \caption{Normalized probability distribution function (PDF) of mean stellar densities, defined as $\rho_{\rm cl} \equiv 3M_{\rm cl}/(8\pi R_{\rm h}^3)$ ({\it blue solid line}).  We display star clusters identified in 31 snapshots coarsely spaced between $z=7$ and $z=5$ (e.g., $\Delta\, t_{\rm out} \sim 17$ Myr for $6< z <7$; sub-sample of clusters shown in Figure \ref{fig:M_vs_Rh} to keep a roughly constant $\Delta\, t_{\rm out}$ between snapshots; see footnote \ref{output-strategy}).  Also shown are the PDFs of gas densities at $z=6.99$ ({\it black dotted line}) and $z=6.92$ (when the cluster ``A'' formed; {\it red dotted line}).  For a reference, the {\it thick dashed vertical line} marks the star formation threshold density.  A bimodal distribution is notable with the majority of mostly unbound associations forming one peak at $\sim 10^{-2}\,\, {\rm M}_{\odot} \,{\rm pc}^{-3}$, and bound clusters forming another peak at $\sim 30 \,\, {\rm M}_{\odot} \,{\rm pc}^{-3}$.  The latter may be a lower limit due to our finite star formation threshold density, but the bimodality is robust.  
\label{fig:rho_vs_N}}
\end{center}
\end{figure}

\begin{figure}
\begin{center}
\includegraphics[width=0.48\textwidth]{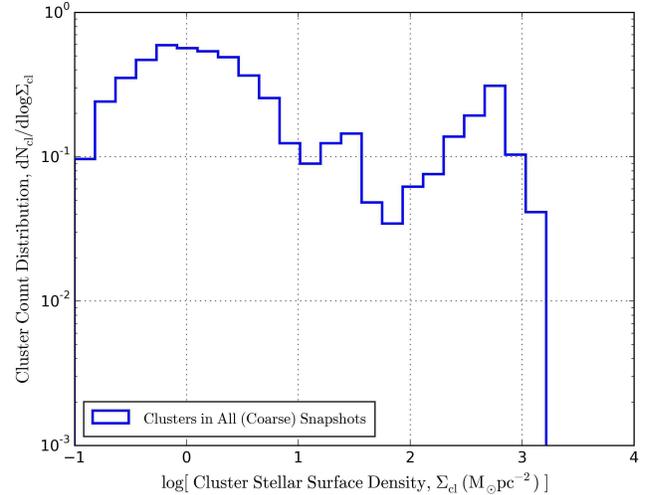}
    \caption{As Figure \ref{fig:rho_vs_N}, but with mean stellar surface densities, defined as $\Sigma_{\rm cl} \equiv M_{\rm cl}/(2\pi R_{\rm h}^2)$.  Interestingly, the broader peak at $\sim 1\,\,{\rm M}_{\odot} \,{\rm pc}^{-2}$ corresponds to what one would obtain by converting a few percent of typical Milky Way giant molecular clouds (GMCs) into stars \protect\citep[e.g.,][]{2009ApJS..181..321E_short}, while the other peak at $\sim 10^3\,\,{\rm M}_{\odot} \,{\rm pc}^{-2}$ corresponds to the surface density identified by \protect\cite{2016arXiv161205635G} as that where feedback begins to become inefficient allowing an order unity fraction of gas to turn into stars (Section \ref{results-formation}).  
    \label{fig:sigma_vs_N}}
\end{center}
\end{figure}

To illustrate the two populations of star clusters, Figure \ref{fig:rho_vs_N} shows the normalized probability distribution function (PDF) of mean cluster stellar densities.  
We see a clear bimodal distribution with the majority of mostly unbound associations forming one peak at $\sim 10^{-2}\,\, {\rm M}_{\odot} \,{\rm pc}^{-3}$, while bound clusters forming another peak at $\sim 30 \,\, {\rm M}_{\odot} \,{\rm pc}^{-3}$.  
The corresponding surface density for this second bimodal group is $\sim 10^{3} \,\, {\rm M}_{\odot} \,{\rm pc}^{-2}$, as can be seen in the surface density PDF of Figure \ref{fig:sigma_vs_N}.  
It is about an order-of-magnitude larger than that of a typical GMC ($\sim 10^2 \,\, {\rm M}_{\odot} \,{\rm pc}^{-2}$), implying a rather unusual dynamical process by which these bound clusters may have formed and evolved.    
The bound cluster's surface density also corresponds to the surface density scale at which star formation over Myr timescales may become highly efficient despite strong stellar feedback \citep[see Section \ref{results-formation} and][]{2016arXiv161205635G}. 
In fact, most of these dense objects can be traced through multiple snapshots and are ancestors of just a few bound clusters (see Section \ref{results-evolution}). 
Note that the second bimodal peak around $\sim 30 \,\, {\rm M}_{\odot} \,{\rm pc}^{-3}$ in Figure \ref{fig:rho_vs_N} corresponds well with the maximum gas density when the cluster ``A'' formed (red dotted line at $z=6.92$).  
This high gas density was realized when two proto-galaxies merge as seen in the second column of Figure \ref{fig:composite_2kpc}, but typically not before or after the merger (black dotted line in Figure \ref{fig:rho_vs_N} at $z=6.99$; representative gas density PDF for a ``normal'' galaxy).  
We will come back to this discussion in Section \ref{results-formation}. 

Figures \ref{fig:rho_vs_N} and \ref{fig:sigma_vs_N} indicate that these bound clusters are in a separate group of a distinctive evolutionary process, not simply a tail of a single, unimodal distribution.   
Of course, because we adopted a finite threshold density for star formation, a patch of gas could have collapsed further -- if we had infinite resolution -- before forming stars.  
So the densities of the bound clusters seen here are probably lower limits.  
This only strengthens our inference of a bimodal population.

\subsection{Evolution of A Long-lasting Bound Star Cluster}\label{results-evolution}

\subsubsection{Method and Overview}\label{results-evolution-1}

We now focus on how the long-lasting bound star clusters evolve in time.
Before moving to discuss the simulation result, a remark on numerical resolution and the ``intra-cluster'' dynamics would be timely.  
The long-lasting bound clusters such as cluster ``A'' are resolved in mass (with $\gtrsim 10^3$ particles), and the force resolution is also well below their half-mass radii. 
Therefore the kinematics amongst the {\it cluster member star particles} is numerically resolved. 
However, this does not mean that our simulation reproduces the phase-space distribution of the {\it actual stars} in observed clusters.  
Partly due to the mass resolution -- each particle represents a stellar population of $\sim 10^3\,{\rm M}_{\odot}$ -- we simply cannot resolve a variety of internal, ``intra-cluster'' evolution processes such as stellar two-body relaxation, mass segregation, evaporation, stellar mass loss, or binary interactions \citep[for a comprehensive review, see e.g.,][]{lrr-2006-2, 2010ARA&A..48..431P}.
Thus, it would be prudent to assume that the reported simulation at best marginally resolves the detailed intra-cluster evolution.  
With this caveat in mind, in all subsequent analyses, we focus {\it only} on globally averaged characteristics of a star cluster, but {\it not} on e.g., radial profiles of stellar density or metallicity inside a cluster, which are more severely affected by the intra-cluster evolution.\footnote{Even some of these averaged characteristics of a simulated cluster should be treated with caution.  For example, the mean densities $\rho$ of long-lasting bound clusters seem to eventually settle in at a value only slightly above the star formation threshold (Figure \ref{fig:progenitor_density}).  Meanwhile their sizes, $R_{\rm h}$, seemingly asymptote to $\gtrsim 10$ pc.  But in nature, obviously, even the massive clusters simulated in this study may become much denser, and more compact in size \citep[see e.g.,][]{2004A&A...416..537L, 2015MNRAS.452..525R}.  The exact values of the cluster's density and size, and its evolution and survival, may still likely depend on the choice of numerical resolution. \label{final-density}}

\begin{figure}
\begin{center}
\hspace*{-0.2cm}\includegraphics[width=0.5\textwidth]{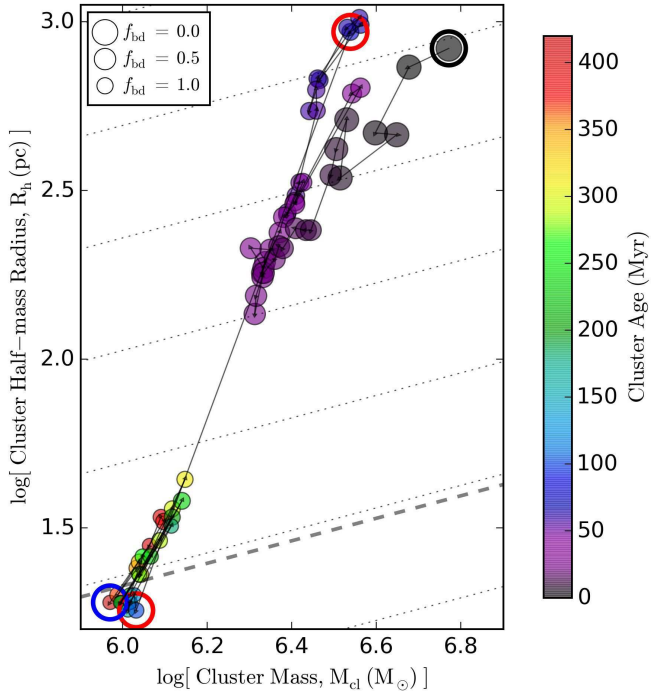}
    \caption{As Figure \ref{fig:M_vs_Rh}, but focusing on the evolution of the cluster ``A'' in a sub-region (gray dashed rectangular region in Figure \ref{fig:M_vs_Rh}).  The {\it thin black line} shows the lineage tree of the cluster ``A'' determined by {\sc Rockstar}.  Each data point is colored by the cluster's age -- different from Figure \ref{fig:M_vs_Rh} -- with its size inversely proportional to its bound mass fraction $f_{\rm bd}$.  The ancestor of the cluster ``A'' begins in the upper right corner of this figure at $z=6.92$ ({\it black circle}), then slowly moves towards the lower left corner to become compact and tightly-bound at $z=5$ ({\it blue circle}).  Initially, a large fraction of the ``association'' mass is due to nearby pre-existing stars and a low-density ``envelope'' of stars, giving an apparently large cluster size.  They are stripped in a mass loss event between  $z=6.4$ and $z=6.3$ (two {\it red circles}), leaving a dense, tightly-bound ``cluster'' that survives for $\sim$420 Myr until the simulation ends.  
\label{fig:M_vs_Rh_zoom-in}}
\end{center}
\end{figure}

In order to trace star clusters evolving in time, we combine the descendant information in standard {\sc Rockstar} outputs, and supplementary information such as member particle IDs of star clusters selected at $z=5$.  
The selected cluster's ``lineage tree'' -- or the main branch of a merger tree -- is built by linking the ``main ancestors'' determined by {\sc Rockstar}. 
(When a cluster P bequeaths the most particles to a cluster D in the next snapshot, P is defined as D's parent, D as P's descendant.  
D may have many parents, the most massive of which is defined as the main ancestor.)\footnote{Or, one may simply locate an ancestor cluster in each snapshot that contains most member particles of the target cluster at $z=5$. The two approaches, in most cases, generate identical lineage trees. But in peculiar periods of evolution involving e.g., galactic major mergers or tidal shocking/stripping, the two techniques could be used in a complementary way.}
We then draw the lineage tree of the cluster ``A'' in Figure \ref{fig:M_vs_Rh_zoom-in}.   
The ancestor of the cluster ``A'' starts out in the upper right corner, then loses its mass at early times to  dramatically evolve to a compact -- i.e., smaller $R_{\rm h}$ and higher mean density -- and tightly-bound -- i.e., higher bound mass fraction -- cluster at $z=5$.  
The following section describes this evolution in detail.  

\begin{figure}
\begin{center}
\hspace*{-0.1cm}\includegraphics[width=0.48\textwidth]{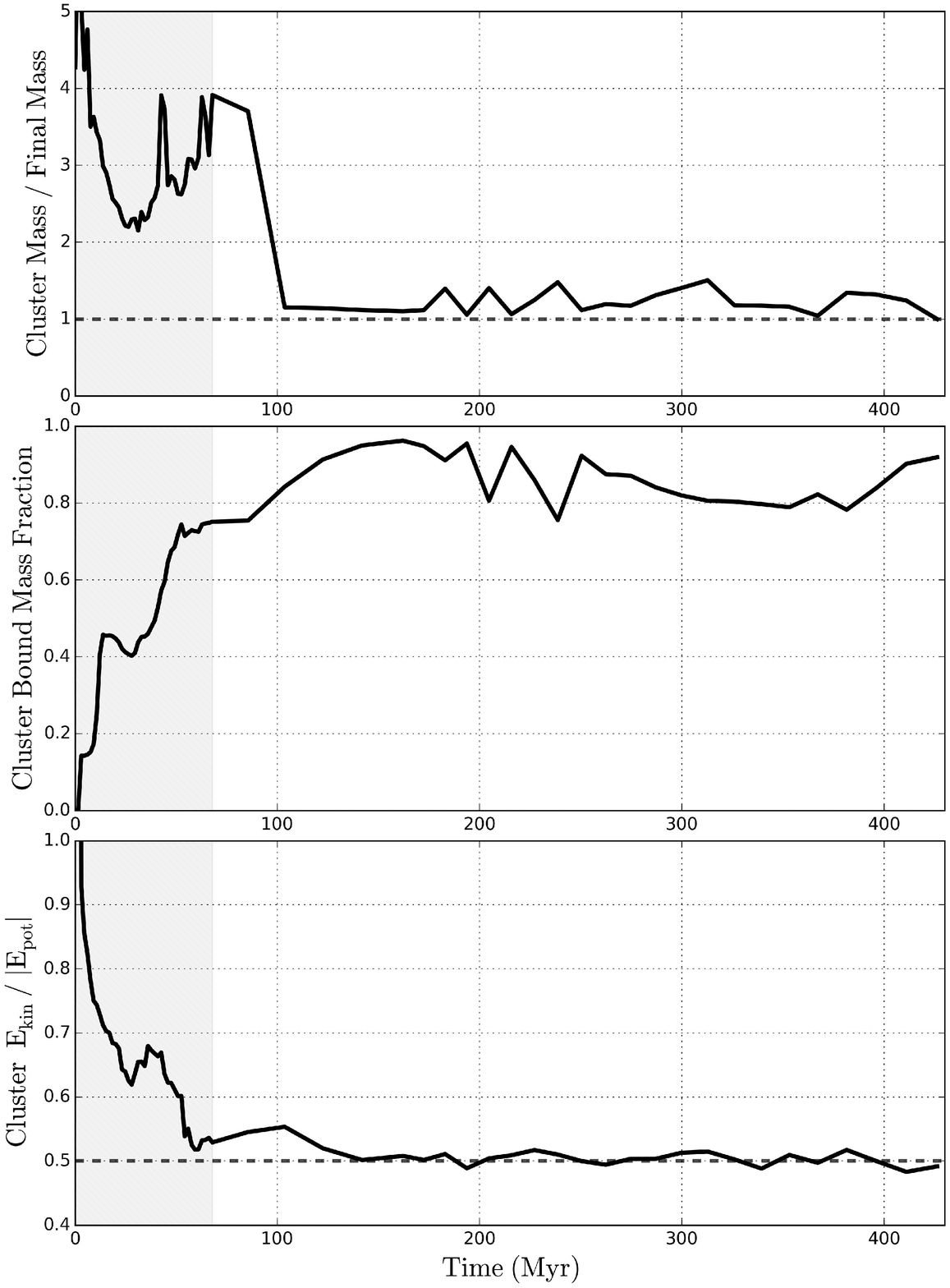}
    \caption{Evolution of the cluster ``A''.  {\it Top:} cluster mass, $M_{\rm cl}$, in units of its final mass at $z=5$ (corresponding to the $x$-axis of Figure \ref{fig:M_vs_Rh_zoom-in}).  The $>$70\% decrease in mass at $\sim$100 Myr refers to the period when the cluster pierces through the host galaxy's gravitational center between  $z=6.4$ and $z=6.3$ (see fifth columns in Figures \ref{fig:star_6kpc}-\ref{fig:composite_2kpc} and Figures \ref{fig:M_vs_Rh_zoom-in}, \ref{fig:progenitor_velocity}). {\it Middle:} cluster bound mass fraction, $f_{\rm bd}$ (corresponding to the color-code of Figure \ref{fig:M_vs_Rh}, or to the inverse of the data point size of Figure \ref{fig:M_vs_Rh_zoom-in}).  The cluster in general becomes more tightly-bound gradually in time.  {\it Bottom:} cluster virial ratio, ${\rm E}_{\rm kin} / |{\rm E}_{\rm pot}| $.  The cluster asymptotes to a dashed horizontal line denoting a virial equilibrium.  0 Myr refers to the moment the cluster ``A'' forms.  The gray shaded region shows the $\sim$80 Myr  period during which the simulation produced outputs at finer intervals (Section \ref{methodology}).  
\label{fig:progenitor_mass}}
\end{center}
\end{figure}

\subsubsection{Three Phases of Cluster Evolution}\label{results-evolution-2}

\begin{figure}
\begin{center}
\hspace*{-0.1cm}\includegraphics[width=0.48\textwidth]{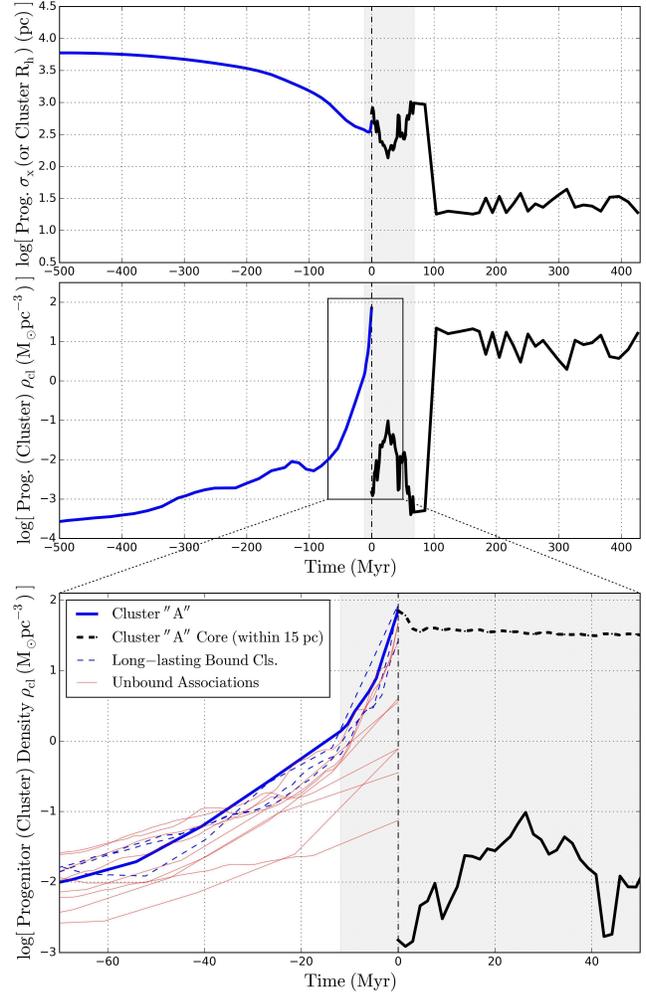}
    \caption{Evolution of the cluster ``A'' and its gas progenitor.  {\it Top:} the gas progenitor member particles' dispersion in position, $\sigma$ ({\it blue}), or the cluster's half-mass radius, $R_{\rm h}$ ({\it black}; corresponding to the $y$-axis of Figure \ref{fig:M_vs_Rh_zoom-in}).  A vertical dot-dashed line in each panel marks 0 Myr, the moment the cluster forms.  The gray shaded region shows the period in which the simulation produced outputs at finer intervals.  {\it Middle:} the gas progenitor member particles' mean density, $\overline{\rho}_{\rm gas}$ ({\it blue}), or the cluster's mean stellar density,  $\rho_{\rm cl} \equiv 3M_{\rm cl}/(8\pi R_{\rm h}^3)$ ({\it black}).  Triggered by a galaxy merger, the gas progenitor's density increases in a short time, causing the gas cloud to collapse and turn a significant portion of its mass, $\gtrsim 10^{5-6}\,{\rm M}_{\odot}$, simultaneously into stars. {\it Bottom:} a zoom-in region around the cluster formation time.  The discontinuity in densities between the blue and black solid lines at 0 Myr is because the newly-formed star cluster grouped by {\sc Rockstar} includes not only the just born star particles but also nearby pre-existing stars and the low-density ``envelope'' of stars forming from nearby less dense gas.  But the gas progenitor's density smoothly transitions to the cluster's core density within 15 pc from its center ({\it thick black dashed line}) that contains almost exclusively newly-born star particles.  In addition to the gas progenitor of the cluster ``A'' ({\it thick blue line}), progenitors of three other long-lasting bound clusters identified at $z=5$ are shown ({\it blue dashed lines}).  Also shown are the progenitors of nine unbound associations at $z=5$ -- that are more massive than $1.5\times 10^5 \,{\rm M}_{\odot}$ at $z=5$, but did not become long-lasting bound clusters ({\it thin red lines}).  All the bound clusters follow a similar evolutionary path, with a large mass collectively reaching high densities in a short timescale ($\sim$10 Myr).  Unbound associations typically form from fragmenting sub-regions within lower-mean-density GMC complexes.  
\label{fig:progenitor_density}}
\end{center}
\end{figure}

For a more quantitative evaluation of the simulation results, evolution of the cluster's various characteristics are shown in thick black solid lines in Figures \ref{fig:progenitor_mass}-\ref{fig:progenitor_velocity_disp}.
For example, Figure \ref{fig:progenitor_mass} depicts the cluster's mass $M_{\rm cl}$, bound mass fraction $f_{\rm bd}$, and virial ratio ${\rm E}_{\rm kin} / |{\rm E}_{\rm pot}|$.  
As in other figures, 0 Myr corresponds to the moment the cluster ``A'' forms.  
The gray shaded region represents the $\sim$80 Myr period during which the simulation produced outputs at finer intervals (Section \ref{methodology}).  

From Figures \ref{fig:M_vs_Rh_zoom-in}-\ref{fig:progenitor_velocity_disp}, three phases of evolution are noticeable: 

{\it (1) Phase 1: Formation and initial evolution of the cluster.}
In the first $\sim$100 Myr  of its life (blue to first red circle in Figure \ref{fig:M_vs_Rh_zoom-in}), the cluster's mass and radius frequently change (Figures \ref{fig:progenitor_mass}-\ref{fig:progenitor_density}), partly because of its proximity to its host galaxy and neighboring clusters.  
In other words, according to our lineage tree analysis  in the first $\sim$100 Myr, the cluster's mass and size change because it loses its loosely-bound member particles, or it captures neighboring particles as it passes by them (but negligible {\it in situ} star formation; see Section \ref{results-composition}).   
The cluster gradually becomes tightly-bound with $f_{\rm bd}$ increasing from $<$10\% to $\sim$80\%, while its virial ratio asymptotes to a virial equilibrium (Figure \ref{fig:progenitor_mass}).\footnote{Because of the relatively low $f_{\rm bd}$ one should be careful when comparing $M_{\rm cl}$ of the objects studied in Figures \ref{fig:M_vs_Rh}, \ref{fig:M_vs_Rh_zoom-in} and \ref{fig:progenitor_mass} with the masses of observed low-redshift GCs for which unbound stars are not included in the calculation of their masses.  For example, few datapoints in Figure \ref{fig:M_vs_Rh} are compatible with the observed GCs since many of them are unbound.  Even the ancestors of the cluster ``A'' in Figure \ref{fig:M_vs_Rh_zoom-in} cannot be directly compared with observed GCs as they are yet to be fully tightly-bound.  Only after the cluster becomes tightly-bound with $f_{\rm bd} \sim 90\%$ can their masses be thought of as consistent with the {\it bound} masses measured for observed GCs.}  
However, as the bottom panel of Figure \ref{fig:progenitor_density} demonstrates, the cluster's core is already very dense at its formation.  
The evolution in mass, size and $f_{\rm bd}$ is primarily due to changes among loosely-bound old stars initially associated with the newly-formed cluster.\footnote{We however caution that the description of the cluster's early evolution in Phases 1-2 may be dependent on the specific algorithm and parameters of the cluster finder ({\sc Rockstar} chosen here; Section \ref{methodology}).  Some of the unbound or loosely-bound stars in the ``envelope'' may not be physically associated with the cluster, and might have been easily lost on a longer timescale even without the tidal process described in Phase 2. \label{rockstar-unbound}}

{\it (2) Phase 2: Mass loss of the cluster.} 
At $\sim$100 Myr after the cluster's formation (between  $z=6.4$ and $z=6.3$; two red circles in Figure \ref{fig:M_vs_Rh_zoom-in}), the cluster's mass decreases by $>$70\% (Figure \ref{fig:progenitor_mass}).
This event refers to the period when the cluster pierces through the host galaxy's gravitational center (fifth columns in Figures \ref{fig:star_6kpc}-\ref{fig:composite_2kpc}), a pericentric approach to the galactic center since the cluster's formation with a very high relative velocity (Figure \ref{fig:progenitor_velocity}).  
Strong tidal shocking at the pericenter and tidal interaction with the host galaxy removes a large fraction of the cluster's mass in a short time, leaving the most tightly-bound core of the clusters, and bringing its bound mass fraction from $\sim$80\%  to $>$90\% (Figure \ref{fig:progenitor_mass}).  
It showcases that the energy gain by a tidal shock can alter the cluster's binding energy \citep[e.g.,][]{2011MNRAS.414.1339K}.
This process preferentially strips the old star particles which existed before the formation of the cluster itself, but were associated with the cluster by {\sc Rockstar} when the cluster formed.  
The process thus leaves the core of tightly-bound young stars with narrower distribution in particle ages  (Sections \ref{results-formation} and \ref{results-composition}).
We speculate that a tidal shock-induced process like this may be essential to transform a population of star clusters to leave only tightly-bound clusters \citep[e.g.,][]{2015MNRAS.454.1658K}.
It is worth noting that the cluster's evolution is driven not only by dynamic relaxation effects of the cluster itself, but also by the embedding galactic contexts such as the cluster passing through strong tidal fields and experiencing mass losses.   

{\it (3) Phase 3: Settling of the cluster.} 
After the mass loss event, the cluster eventually settles in at $M_{\rm cl} \sim 10^{6.0}\, {\rm M}_{\odot}$ and $R_{\rm h} \sim 10^{1.4}\,{\rm pc}$.  
The cluster maintains these properties for the next $\sim$320 Myr.   
It also preserves most other characteristics such as the bound mass fraction, virial ratio, density, metallicity, and velocity dispersion (Figures \ref{fig:progenitor_mass}-\ref{fig:progenitor_density} and \ref{fig:progenitor_velocity_disp}). 
One of the reasons that the cluster could maintain its characteristics is because the cluster has now moved to an orbit with a relatively large radius after a fast velocity increase around the major mass loss event in Phase 2 (Figure \ref{fig:progenitor_velocity}).  
Our observation is broadly consistent with \cite{2015MNRAS.454.1658K} who argued that external perturbation events such as galaxy mergers could cause clusters to migrate into the halo, and thus limit the duration of the shock-induced disruption phase.    
The cluster may survive for an extended period of time since it is now less likely to undergo tidal shocks that could disrupt it.
It is also interesting that the cluster is ejected from the main concentration of stars into the halo, where Milky Way's metal-poor ``blue'' GCs are indeed found.\footnote{Clearly, the limited run time of our simulation prohibits us from predicting the ultimate fate of our simulated high-redshift clusters.  An extensive study with more samples would be needed to thoroughly test this hypothesis.  For recent studies closely related to this idea, see e.g., \cite{2015MNRAS.454.1658K} and \cite{2015arXiv151108212A}.}

\subsection{Formation of A Long-lasting Bound Star Cluster}\label{results-formation}

\subsubsection{Method and Overview}\label{results-formation-1}

\begin{figure}
\begin{center}
\hspace*{-0.1cm}\includegraphics[width=0.48\textwidth]{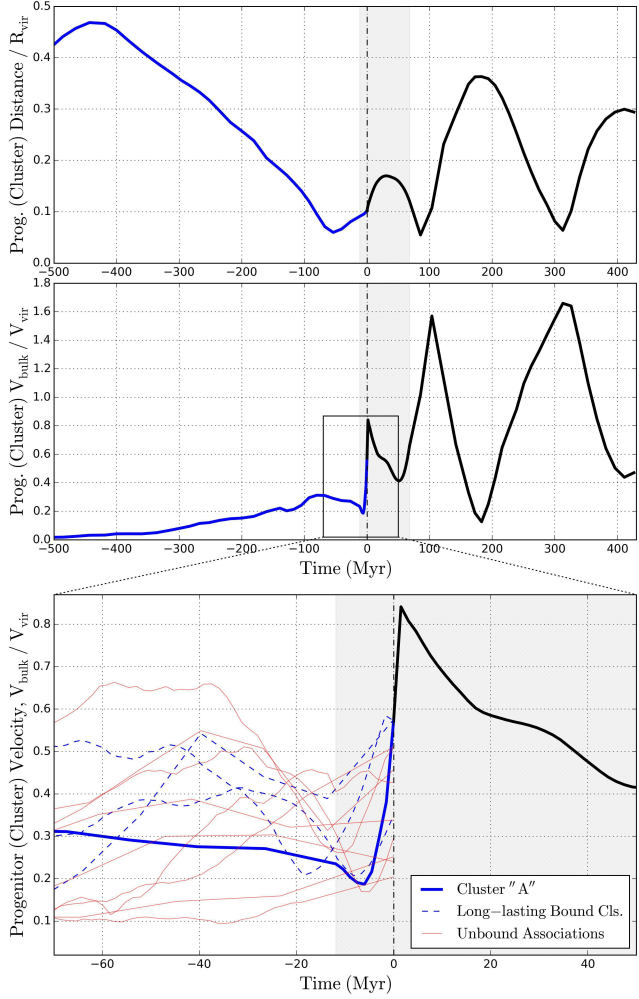}
    \caption{Evolution of the cluster ``A'' and its gas progenitor as  Figure \ref{fig:progenitor_density}, but in different properties.  {\it Top:} gas progenitor member particles' ({\it blue}) or the cluster member particles' ({\it black}) mean distance to the galactic center normalized by the galactic virial radius $R_{\rm vir,\, gal}$.  {\it Middle:} gas progenitor member particles'  ({\it blue}) or the cluster member particles' ({\it black}) bulk velocity with respect to the galactic center  normalized by the galactic virial velocity $V_{\rm vir, \,gal}$.  {\it Bottom:} a zoom-in region around the formation time of the cluster ``A''.   The cluster formation event is associated with a drastic velocity change around a pericentric passage during a proto-galaxy merger (i.e., strong tidal shock with Mach number $\sim$100; see Figure \ref{fig:composite_2kpc}).
\label{fig:progenitor_velocity}}
\end{center}
\end{figure}

\begin{figure}
\begin{center}
\hspace*{-0.12cm}\includegraphics[width=0.486\textwidth]{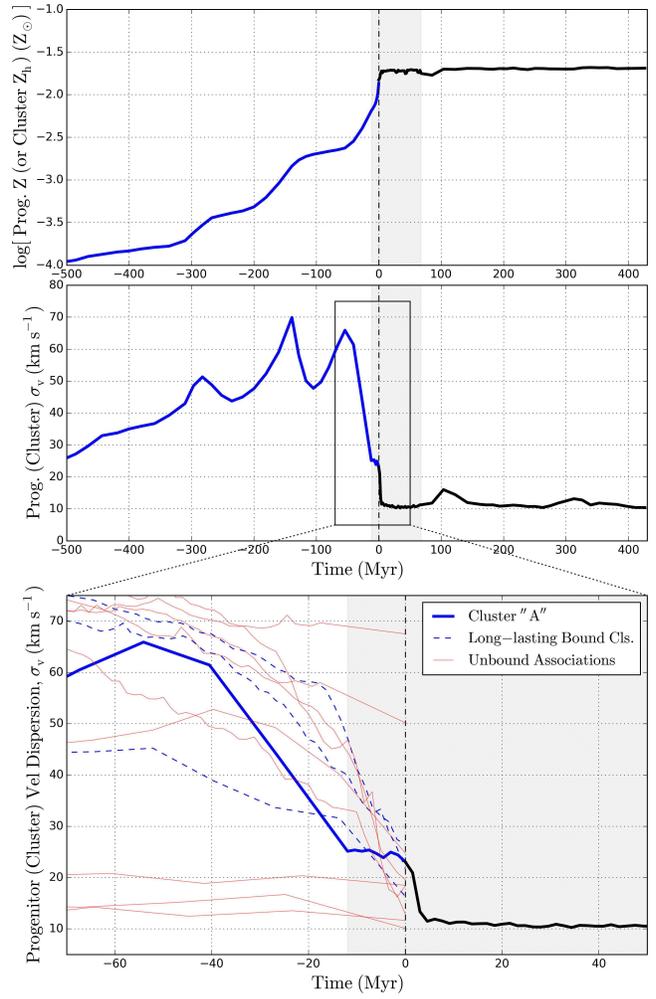}
    \caption{Evolution of the cluster ``A'' and its gas progenitor as Figure \ref{fig:progenitor_density}, but in different properties.  {\it Top:} gas progenitor member particles' ({\it blue}) or the cluster's ({\it black}) mean metallicity within a half-mass radius, $Z_{\rm h}$.  {\it Middle:} gas progenitor member particles' ({\it blue}) or the cluster member particles' ({\it black}) velocity dispersion, ${\sigma}_{\rm v}$.  {\it Bottom:} a zoom-in region around the formation time of the cluster ``A''.  The sharp velocity change at the cluster formation time (Figure \ref{fig:progenitor_velocity}) is associated with a sharp drop in velocity dispersion, consistent with the tidal compression.  
\label{fig:progenitor_velocity_disp}}
\end{center}
\end{figure}

In this section, we go back to the moment when these long-lasting bound star clusters form, and examine the formation conditions of the bound clusters.  
In particle-based hydrodynamics codes like {\sc Gizmo} a newly-born star particle inherits the unique particle ID number from its progenitor gas particle.  
Using these unique IDs, we can identify the ``gas progenitor'' of a cluster in an earlier snapshot before the cluster formed.  
The cluster's ``formation time'' is defined as the moment when 50\% of the gas progenitor member particles have turned into star particles.  
Using such information, in Figures \ref{fig:progenitor_density} to \ref{fig:progenitor_velocity_disp} we present the evolution of the cluster ``A'' and its gas progenitor.
In addition to the gas progenitor of the cluster ``A'', progenitors of three other long-lasting bound clusters identified at $z=5$ are shown (blue dashed lines in the bottom panels; recall that these three clusters are more massive than  $10^{5.5}\,\,{\rm M}_{\odot}$ and survived $>$ 300 Myr at $z=5$; marked by thin blue circles in Figure \ref{fig:M_vs_Rh}).
Also shown are the progenitors of nine unbound associations at $z=5$ -- that are more massive than $1.5\times 10^5 \,{\rm M}_{\odot}$ at $z=5$, but fail to meet the criteria of a long-lasting bound cluster (thin red lines).

At early times, more than 100 Myr before the formation, the gas progenitor particles of the cluster ``A'' are spread over $\gtrsim$ kpc, effectively a random sample of the entire galaxy (Figure \ref{fig:progenitor_density}).  
About 30 Myr before the formation time, the gas progenitor's mean density reaches $\sim 10\,\, {\rm cm}^{-3}$ (or $\sim 10^{-1}\,\,{\rm M}_{\odot} \,{\rm pc}^{-3}$), a typical value of a dense ISM in this relatively dense high-redshift galaxy.    
Then for the next 30 Myr, a galaxy merger event disturbs the gravitational potential around the gas progenitor sitting between the two merging proto-galaxies (first and second  columns in Figures \ref{fig:star_6kpc}-\ref{fig:composite_2kpc}).  
This event forces the gas cloud into strong compressive shocks (Figure \ref{fig:progenitor_velocity}), forming a self-gravitating cloud with its velocity dispersion decreased to $\lesssim 20\,\,{\rm km\,s}^{-1}$ (Figure \ref{fig:progenitor_velocity_disp}).  
The {\it mean} density of the gas progenitor accordingly increases quickly to $\gtrsim 10^3\,\, {\rm cm}^{-3}$ (or $\gtrsim 10\,\,{\rm M}_{\odot} \,{\rm pc}^{-3}$; see Figure \ref{fig:progenitor_density}, or the red dotted line of Figure \ref{fig:rho_vs_N}).
Consequently, the gas cloud collapses and turns a significant portion of its mass ($\gtrsim 10^{5-6}\,{\rm M}_{\odot}$, comparable to that of a star cluster) simultaneously into stars in the gas free fall timescale, $t_{\rm ff} \lesssim 3$ Myr (see also Section \ref{results-composition}).  
Due to its high density and compactness, the newly-formed cluster remains gravitationally bound.    
Note that this formation scenario applies also to other long-lasting bound star clusters.

In order to illustrate the nature of gas compression triggering star cluster formation, in Figure \ref{fig:radial_velocities} we plot the radial profile of the mass-weighted gas infall velocity when the cluster ``A'' starts to form.
The plot is centered on the gas progenitor of the cluster -- the location of densest gas within 100 pc from the center of the gas progenitor -- at 10.5 Myr before the cluster's formation time (when the first stars in the cluster begin to form).  
The negative velocity values indicate the inward motion of the gas.
The merger event disturbs the gravitational potential around the progenitor and compresses the gas cloud to high density where it is strongly self-gravitating.
Therefore, the actual gas infall speed is of the order of the purely gravitational infall speed at each radius $R$, $v_{\rm ff} (R) = \sqrt{8GM_{\rm tot,\,<}(R)/(\pi^2R)}$, where $M_{\rm tot,\,<}(R)$ is the total mass within $R$.  
The gas infall speed does occasionally become larger than the purely gravitational collapse speed.  
This is indicative of a massive gas cloud with little support from rotation or shear, that collapses under self-gravity until it undergoes efficient star formation at high density.

\subsubsection{Merger-induced Star Cluster Formation}\label{results-formation-2}

\begin{figure}
\begin{center}
\includegraphics[width=0.45\textwidth]{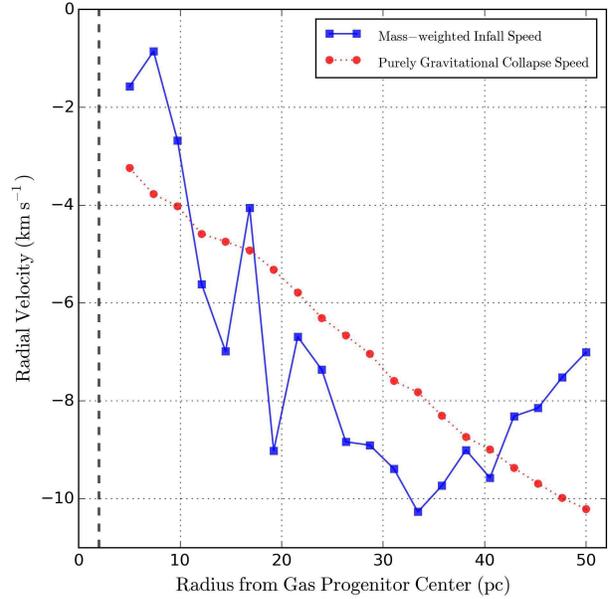}
    \caption{Radial profile of the mass-weighted gas infall velocity around the gas progenitor of the cluster ``A'' when it begins to form ({\it blue solid line}; 10.5 Myr before the cluster's formation time).  The negative values indicate the inward movement of the gas.  Also shown is the purely gravitational infall speed at each radius ({\it red dotted line}).  The {\it thick dashed vertical line} marks the minimum force resolution for gas.  The galaxy merger event disturbs the gravitational potential around the progenitor and compresses the gas cloud, making the actual gas infall speed occasionally larger than the purely gravitational collapse speed.  For cold, molecular star-forming gas, the compressive velocities here corresponds to Mach $\sim$ 50 flows.  
\label{fig:radial_velocities}}
\end{center}
\end{figure}

Overall, our investigation paints a picture of bound star cluster formation when massive clouds reach very high densities -- a condition which may be preferentially {\it merger-induced} (though in principle similar conditions could occur without mergers). 
Mergers can produce tidal shocks in regions of strong tidal acceleration (e.g., bridges, tidal tails, induced bars) with velocity jumps of order the circular velocity (Figure \ref{fig:progenitor_velocity}), corresponding to Mach numbers of $\sim 50-100$, which in turn produces extreme gas compression \citep[e.g.,][]{2009ApJ...694L.123K}. 
The resulting dense gas clumps efficiently cool and collapse.  
The free fall time at the velocities shown in Figure \ref{fig:radial_velocities} reaches $\sim 0.5-1.5$ Myr for the clump core with densities $\sim 10^{3-4}\,{\rm cm}^{-3}$. 
Therefore collapse occurs {\it before} the first SNe explode (which requires a time $\sim 3-10$ Myr).
Although radiative feedback and stellar winds are present in our simulation and act on shorter timescales, the gas densities are sufficiently high such that they are unable to disrupt the cloud until a significant fraction -- an order unity -- of the gas has been converted into stellar mass \citep{2016arXiv161205635G}.\footnote{For a cloud core with $\sim$$10^{6} \,\,{\rm M}_{\odot}$ inside $\sim$10 pc (Figure \ref{fig:M_vs_Rh_zoom-in}), the escape velocity is $\gtrsim 20\,\,{\rm km\,s}^{-1}$, so photoionization heating which heats the gas to the thermal speed of only $\sim 10\,\,{\rm km\,s}^{-1}$ cannot unbind the cloud. Radiation pressure and stellar winds for a zero-age main sequence population both carry a momentum flux $\sim L/c \sim 10^{32}\,\,{\rm dyne}\,(M_{\star}/10^{6}\,\,{\rm M}_{\odot})$, which is less than the gravity $\sim G\,M_{\rm cl}\,M_{\rm gas}/R_{\rm h}^{2} \sim 10^{32}\,\,{\rm dyne}\,(M_{\rm gas}/10^{6}\,\,{\rm M}_{\odot})\,(\Sigma_{\rm cl}/10^3\,\,{\rm M}_{\odot}\,{\rm pc^{-2}})$ for a clump surface density $\gtrsim 10^3\,\,{\rm M}_{\odot}\,{\rm pc^{-2}}$, until most of the gas is turned into stars (see \citealt{2016arXiv161205635G} for more details).} 
The resulting, highly efficient star formation allows the cluster to remain bound. 
This is in line with previous analytic \citep{2010ApJ...710L.142F, 2012MNRAS.426.3008K} and numerical \citep{2016arXiv161205635G} models for the fraction of stars that form in bound clusters. 

By disturbing and compressing the system, galaxy mergers provide unique opportunities to make the otherwise normal ISM shock to abnormally high densities.
The violent environment is important to push a large gas mass of $\gtrsim 10^{5-6}\,{\rm M}_{\odot}$ {\it collectively} to very high density in a short time, rather than letting it fragment into smaller dense pieces.
Once at such high density, the cloud promptly turns a significant portion of itself into stars at a small dynamical timescale, $t_{\rm ff} \lesssim 3$ Myr.  
Since proto-galaxies at high redshift frequently experience mergers, they could have provided a fertile environment to produce multiple long-lasting bound clusters -- which we speculate to be candidates for present-day ``blue'' GCs.   

We emphasize the critical difference between our merger-induced formation process of long-lasting bound star clusters and the {\it normal} turbulent fragmentation.  
All star particles require a density above $n_{\rm th} = 500 \,\,{\rm cm}^{-3}$ to spawn in our simulation (Section \ref{methodology}), yet most of them end up in unbound associations.  
It is because, in a typical star-forming gas clump, only a small fraction of its mass in sub-clumps (with mass $\ll 10^{5}\,{\rm M}_{\odot}$) is sufficiently dense to form star particles.  
The newly-born star particles then destroy their parent cloud complex.
By contrast, in the cluster formation scenario identified in our simulation, a violent galaxy merger causes a  large amount of self-gravitating gas mass -- comparable to that of a star cluster -- to simultaneously reach high density above the star formation threshold.  

\subsubsection{Caveats}\label{results-formation-3}

It is worth discussing a few points about this process:  

{\it (1)} Since the newly-formed star cluster grouped by {\sc Rockstar} includes not only the just born star particles but also nearby pre-existing ones, the newly-formed cluster's mean stellar density could be smaller than that of the gas progenitor of the cluster ``A'' at $z=5$.  
This results in the discontinuity in densities between the blue and black solid lines at 0 Myr in the middle/bottom panels of Figure \ref{fig:progenitor_density}.  
In fact, at $z=6.90$, just $\sim$3.2 Myr after its formation, the cluster ``A'' is composed of $\sim$80\% in pre-existing stars, and only $\sim$20\% in stars just born in the past 5 Myr (see Section \ref{results-composition} for more discussion).  
The pre-existing stars are typically in the outskirts of the cluster, and tend to be preferentially removed already by $z=5$, leaving only the core of tightly-bound young stars (but again, see footnote \getrefnumber{rockstar-unbound}).
But notice that if we consider the mean density of the core within 15 pc containing almost exclusively newly-born star particles, the gas progenitor's density smoothly transitions to the newly-formed cluster's core density at 0 Myr (thick black dashed line in the bottom panel of Figure \ref{fig:progenitor_density}).  

{\it (2)} Readers should note that the high-density cluster formation process described above is not a {\it sufficient} condition, but only a {\it necessary} condition for a long-lasting bound cluster.  
Among the nine unbound associations shown in Figures \ref{fig:progenitor_density}-\ref{fig:progenitor_velocity_disp} (thin red lines), three of them broadly followed similar paths as that of the cluster ``A'' when formed.
But eventually, they do not become long-lasting bound clusters at $z=5$.   
Other processes after their formation -- such as tidal disruption or gravitational capture -- interrupt their lives as bound clusters at some points before $z=5$.  
Obviously, however, we cannot exclude the possibility that they could have survived as bound clusters if better numerical resolution had been adopted.

\begin{figure}
\begin{center}
\hspace*{-0.1cm}\includegraphics[width=0.48\textwidth]{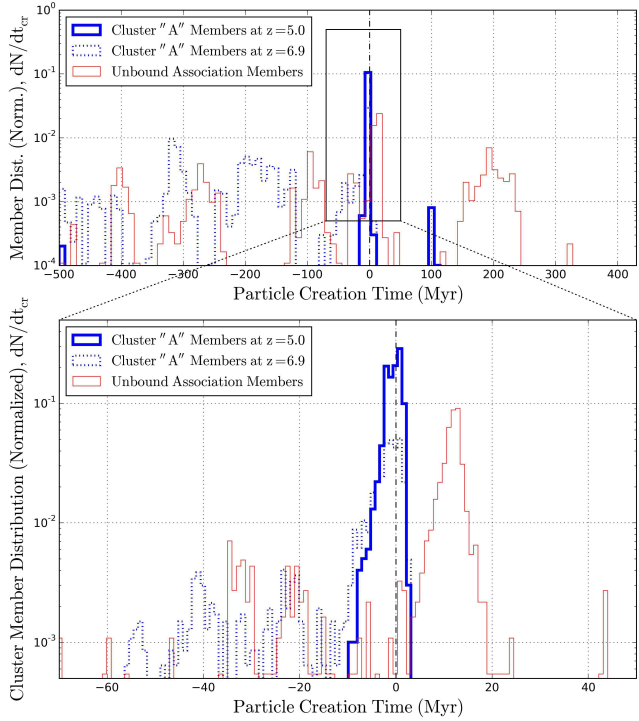}
    \caption{{\it Top:} The normalized PDF of the star cluster member particles' creation times $t_{\rm cr}$ (time bin size $=$ 9.3 Myr).  0 Myr corresponds to the moment each star cluster forms.  A {\it thick blue line} and a {\it blue dotted line} are for the members of the cluster ``A'' at $z=5$ and $z=6.90$, respectively, while a {\it thin red line} is for a representative unbound association (see Section \ref{results-composition} for how this unbound association is selected).  {\it Bottom:} a zoom-in region around the formation time of the cluster ``A'' with finer binning in time (bin size $=$ 0.93 Myr).  The $x$-axis range is kept identical among Figures \ref{fig:progenitor_mass}-\ref{fig:progenitor_velocity_disp} and \ref{fig:progenitor_creation_time} for easier comparison.  Most of the cluster's member particles left by $z=5$ were formed in a short burst, within $\Delta\, t_{\rm cr} \sim t_{\rm ff} \lesssim 3$ Myr from the cluster's formation time.  
\label{fig:progenitor_creation_time}}
\end{center}
\end{figure}

\begin{figure}
\begin{center}
\includegraphics[width=0.48\textwidth]{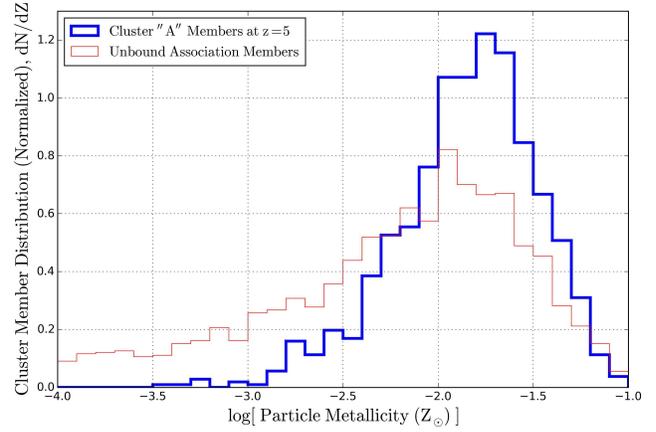}
    \caption{The normalized PDF of the star cluster member particles' metallicities.  The {\it thick blue line} is for the cluster ``A'' at $z=5$, while the {\it thin red line} is for an unbound association (the same one as in Figure \ref{fig:progenitor_creation_time}).  In this simulation, gas particles have fixed metallicity with no inter-particle mixing allowed.  The metallicity spread we find should be interpreted as an upper limit of the true metallicity spread -- if e.g., turbulence mixes metals on small scales in star cluster-forming regions.  
\label{fig:progenitor_metallicity}}
\end{center}
\end{figure}

It is also useful to comment on the numerical aspects of our simulation:

{\it (1)} As pointed out before (footnote \getrefnumber{final-density}), what may {\it not} be so robust in our calculation is the quantities like the cluster's final size and density, because the cluster's relaxation effects are only marginally resolved. 
In contrast, what {\it is} indeed physically robust is the fact that a large self-gravitating gas mass of $\gtrsim 10^{5-6}\,{\rm M}_{\odot}$ could reach very high density {\it simultaneously} in galaxy mergers, and should very efficiently form stars before any of the various stellar feedback channels intervenes. 
Based on our arguments above, capturing this in simulations requires the ability to resolve surface densities at least as high as $\sim 10^3\,\,{\rm M}_{\odot} \,{\rm pc}^{-2}$.
In nature, gas clouds may reach even higher densities before they become self-gravitating and star-forming.

{\it (2)} We argue that a numerical simulation with high resolution and a high dynamic range needs to be accompanied by appropriate subgrid physics to properly describe the clustered star formation scenario described here.  
For example, we find it harder to identify long-lasting bound clusters in the same runs with lower star formation threshold density \citep[e.g., {\it z5m10mr} run with $n_{\rm th} =  50\,\, {\rm cm}^{-3}$ in][see footnote \ref{ma-threshold}]{2015MNRAS.453..960M} or without a self-gravity star formation criterion \citep[e.g., {\it z5m10e} run in][]{2015MNRAS.453..960M}, because in this cases stars artificially form at such low densities that clumps can never reach the critical mean densities described above (i.e., $\gtrsim 10^3\,\, {\rm cm}^{-3}$).\footnote{This result is in line with Figure 10 of \cite{2017ApJ...834...69L} where only with high enough star formation threshold can they reproduce the short burst of star formation found in observed young massive clusters.}  
This suggests that for the hydrodynamics code adopted in our study, a physically-motivated star formation criterion is crucial to depict the Jeans gravitational collapse at the relevant mass and length resolution scales, and to numerically reproduce extraordinary conditions such as merger-induced star cluster formation.

\subsection{Composition of A Long-lasting Bound Star Cluster At The End of The Simulation ({\textit {\textbf z}} ${\bf =}$ {\textit {\textbf 5}})}\label{results-composition}

Finally we examine the cluster properties at $z=5$ focusing on how the cluster formation event is imprinted in its composition.  
Figure \ref{fig:progenitor_creation_time} shows the normalized PDF of the star cluster member particles' creation times $t_{\rm cr}$.  
A small dispersion in member particles' ages for the cluster ``A'' (clustered at 0 Myr) reinforces the idea that the star formation episode lasted for a very short time.
The two other small peaks located at $-$500 Myr and 100 Myr contribute negligibly to the total cluster mass. 
For comparison, we also plot the $t_{\rm cr}$ distribution for the cluster ``A'' at $z=6.90$, just $\sim$3.2 Myr after its formation.  
Even though the cluster was initially grouped with nearby pre-existing stars of varying ages by {\sc Rockstar}, they are typically in the outskirts of the cluster, and tend to be preferentially removed already by $z=5$. 
This process left only the core of tightly-bound young stars with very narrow distribution in particle ages. 
We also plot the PDF of one of the six unbound associations whose gas progenitors are shown in thin red lines in Figures \ref{fig:progenitor_density}-\ref{fig:progenitor_velocity_disp}, but do not resemble the evolution path of the cluster ``A'' progenitor.
Indeed, its $t_{\rm cr}$ distribution reveals multiple generations of member star particles.

The width of the main spike in the cluster ``A'' members' formation episode is of order the free fall time (bottom panel of Figure \ref{fig:progenitor_creation_time}).  
$\sim$60\% of the entire member star particle population were born within a $\sim$3 Myr period in this main spike.  
That is, $\Delta\, t_{\rm cr} \sim 3\,\, {\rm Myr} \sim t_{\rm ff}$, which is approximately consistent with the recent observations of young massive clusters \citep{2013MNRAS.436.2852B, 2014MNRAS.445..378B, 2014MNRAS.441.2754C}. 

In Figure \ref{fig:progenitor_metallicity} we plot the normalized PDF of the star cluster member particles' metallicities.  
This agrees reasonably well with internal abundance variations found in recent observations of metal-poor ``blue'' GCs \citep[e.g.,][]{2004ARA&A..42..385G, 2009ApJ...705.1481D, 2015MNRAS.450..815M, 2016MNRAS.455.2417R}.    
It also shows that the star-forming gas was already enriched up to as high as $10^{-1}\, Z_{\odot}$, while approximately matching the observed mean abundance value.  
(One may also speculate that the cluster ``A'' exhibits a slightly narrower distribution than a representative unbound association does.)
However, the large spread in the member particle metallicities for the unbound association (larger than the observed [Fe/H] spreads of $\lesssim$ 0.05 dex), or equivalently a large noise in the metallicity field, is to a great extent a numerical artifact because we do not include a sophisticated metal mixing scheme in the simulation -- that is, presently metals are simply locked into individual particles once deposited by SNe.  
We plan to investigate if including a subgrid-scale turbulent metal diffusion scheme \citep[e.g.,][]{2017MNRAS.471..144S} would reduce the artificially enhanced metallicity spread.

\section{SUMMARY AND CONCLUSION}\label{conclusion}

Using a state-of-the-art cosmological simulation of high-redshift merging proto-galaxies from the {\it FIRE} project, we have investigated the formation and evolution of star clusters, and in particular explored one formation hypothesis for present-day metal-poor ``blue'' GCs. 
In the simulation, two populations of clustered star formation emerge. 
Most stars form in unbound or loosely-bound associations with mean stellar densities $\sim 10^{-2}\,\, {\rm M}_{\odot} \,{\rm pc}^{-3}$, or surface densities $\sim 1\,\, {\rm M}_{\odot} \,{\rm pc}^{-2}$, corresponding to a typical GMC complex converting a few percent of its mass into stars. 
This is the behavior we have previously shown as ``typical'' in simulations with similar physics at low redshifts \citep{2012MNRAS.421.3488H} and seen in observations \citep[e.g.,][]{2009ApJS..181..321E_short}. 
However, a few percent of stars form in bound clusters with mean stellar densities $\sim 30 \,\, {\rm M}_{\odot} \,{\rm pc}^{-3}$, or surface densities $\sim 10^3 \,\, {\rm M}_{\odot} \,{\rm pc}^{-2}$ (Section \ref{results-population}). 
At these surface densities, \cite{2016arXiv161205635G} showed (in simulations of individual cluster-forming clouds, at sub-${\rm M}_{\odot}$ resolution but with the same physics) that stellar feedback begins to become unable to efficiently expel the cloud before/as it collapses into stars, leading to an order unity fraction of gas turning into stars. 
Essentially, the self-gravity of the cloud becomes too large for small-scale feedback channels to overcome, particularly as the dynamical times for collapse become shorter than the massive star's evolution timescale -- and shorter than the timescale for SNe explosion. 
A similar threshold was motivated by simple analytic arguments in \cite{2010ApJ...710L.142F}. 
Critically, the simulation here see {\it both} populations, with sufficient dynamic range to show that the same physics which produces low star formation rates in a galaxy-averaged sense \citep{2014MNRAS.445..581H, 2017arXiv170101788O} and most of the star formation in unbound associations can also produce a reasonable mass fraction in dense, bound, GC-like objects. 
It is also crucial that the simulation here include stellar feedback not just from SNe but also from photoheating, radiation pressure and stellar winds, since the cluster formation timescales are shorter than $\lesssim 3$ Myr, so SNe do not have the chance to explode before the clusters form (Section \ref{methodology}).  
In other words, simulations which do not include these other feedback mechanisms could therefore easily overestimate the formation efficiency of dense clusters. 

We show that frequent mergers in high-redshift proto-galaxies can provide a fertile environment for the production of this GC-like population. 
The mergers disturb the gravitational potential and produce tidal shocks with Mach numbers $\sim 50-100$, which rapidly and collectively push (pre-existing) large gas clouds of $\gtrsim 10^{5-6}\,{\rm M}_{\odot}$ to very high densities $> 10^{3}\,{\rm cm}^{-3}$ (Section \ref{results-formation}). 
Such clouds then cool and collapse to form stars extremely rapidly before stellar feedback can stop star formation. 
We typically see a dense ``core'' in these systems, which is tightly-bound, surrounded by a lower-density ``envelope'' of loosely associated stars and nearby pre-existing stars. 
Those stars are quickly and preferentially stripped -- not just by dynamical relaxation effects (which we do not necessarily resolve in the simulations) but also by the cluster passing through the tidal field of its host galaxy (Section \ref{results-evolution}). 
In the simulation we analyzed, the tightly-bound core is left behind and survives for $\sim$420 Myr, until the end of the simulation. 
This relic core has a very small age spread, and relatively small metallicity spread (Section \ref{results-composition}).

Of course, the simulation here still suffers from various limitations. 
Our limited resolution, 800 $h^{-1}{\rm M}_{\odot}$, means that we can only follow very massive clusters, and cannot follow detailed intra-cluster evolution (dynamical relaxation, mass segregation, etc; Section \ref{results-evolution}). 
Next-generation simulations in progress reach $\sim 30\,\,{\rm M}_{\odot}$ resolution (Wheeler et al. in prep.), and in follow-up studies we can use these new runs to identify cluster-forming regions which are then refined to gain superior resolution. 
The limited evolution time of the present simulation -- stopped at $z=5$ -- entails that we cannot yet predict the ultimate fate of the clusters (e.g., their locations in a $z=0$ Milky Way-like galaxy, into which this progenitor halo should eventually evolve). 
Future simulations will also address this (Wetzel et al. in prep.), but also methods which replace clusters formed self-consistently with tracer particles which can be evolved for longer times \citep[e.g.,][]{2017MNRAS.465.3622R, 2017ApJ...834...69L} could be useful.
Improvements to the hydrodynamics and subgrid physics in our new {\it FIRE-2} model \citep{2017arXiv170206148H_short} may also influence our predictions, although preliminary comparisons suggest that the hydrodynamic method (e.g., {\sc Gizmo}'s mesh-free finite-mass mode versus the {\sc P-sph} mode chosen here) does not significantly affect the conclusions presented here. 
Future work will include analyzing different runs performed with different hydrodynamics solvers and  star formation prescriptions to constrain their effects in the stellar distribution and cluster numbers.  
More accurate treatments of stellar yields (as opposed to the IMF-averaged yields adopted here) and subgrid metal diffusion in the dense ISM will be critical to make testable predictions for the internal abundance patterns within GCs, an obvious frontier for observations. 

\vspace{-0.5cm}

\section*{Acknowledgments}

The authors thank Tom Abel, Nathan Bastian, Diederik Kruijssen, Joel Primack and Eros Vanzella for useful discussions and insightful comments during the progress of this study.
Ji-hoon Kim acknowledges support from NASA through an Einstein Postdoctoral Fellowship grant PF4-150147 awarded by the Chandra X-ray Center, which is operated by the Smithsonian Astrophysical Observatory for NASA under contract NAS8-03060. 
He also acknowledges support from the Moore Center for Theoretical Cosmology and Physics at Caltech.  
He thanks for the support from the computational team at SLAC National Accelerator Laboratory, from Shawfeng Dong at the University of California Santa Cruz, and from Chris Mach at Caltech during the usage of the clusters for simulation analyses.  
The computing time used for the presented simulations was provided by Extreme Science and Engineering Discovery Environment (XSEDE) allocations TG-AST120025, TG-AST130039, TG-AST140023 and TG-AST140064.  
XSEDE is supported by the National Science Foundation (NSF) grant ACI-1053575.
Philip Hopkins acknowledges support by the Gordon and Betty Moore Foundation through grant 776 to the Caltech Moore Center for Theoretical Cosmology and Physics, by the Alfred P. Sloan Foundation through Sloan Research Fellowship BR2014-022, by the NSF through grant AST-1411920 and CAREER grant 1455342, and by NASA through the ATP grant NNX14AH35G.  
The Flatiron Institute is supported by the Simons Foundation.
Andrew Wetzel acknowledges support by a Caltech-Carnegie Fellowship, in part through the Moore Center for Theoretical Cosmology and Physics at Caltech, and by NASA through grant HST-GO-14734 from STScI.
Claude-Andr\'{e} Faucher-Gigu\`{e}re acknowledges support by the NSF through grants AST-1412836 and AST-1517491, by NASA through grant NNX15AB22G, and by STScI through grants HST-AR-14293.001-A and HST-GO-14268.022-A. 
Du\v{s}an Kere\v{s} acknowledges support by the NSF through grant AST-1412153, and by the Cottrell Scholar Award from the Research Corporation for Science Advancement.
Shea Garrison-Kimmel acknowledges support from NASA through an Einstein Postdoctoral Fellowship grant PF5-160136. 
The publicly available {\tt yt} code used in the analysis of this work is the product of collaborative efforts by many independent scientists from numerous institutions around the world.  
Their commitment to open science has helped make this work possible.

\label{lastpage}

\end{document}